\documentclass[showpacs,superscriptaddress,twocolumn,tightenlines,amsmath,amssymb]{revtex4}

\usepackage{color}
\usepackage[dvipdfm]{graphicx}

\begin{document}

\title{Dark Matter and Higgs Boson in a Model with Discrete Gauge Symmetry}
\author{Cheng-Wei Chiang}  
\affiliation{Department of Physics and Center for Mathematics and Theoretical Physics,
National Central University, Chungli, Taiwan 32001, Republic of China\smallskip}
\affiliation{Institute of Physics, Academia Sinica, Taipei, Taiwan 11529, Republic of China\smallskip}
\affiliation{Physics Division, National Center for Theoretical Sciences, Hsinchu, Taiwan 30013, Republic of China\smallskip}
\author{Takaaki Nomura}  
\affiliation{Department of Physics and Center for Mathematics and Theoretical Physics,
National Central University, Chungli, Taiwan 32001, Republic of China\smallskip}
\author{Jusak Tandean}  
\affiliation{Department of Physics and Center for Mathematics and Theoretical Physics,
National Central University, Chungli, Taiwan 32001, Republic of China\smallskip}
\affiliation{Department of Physics and Center for Theoretical  Sciences, National Taiwan University,\\
Taipei, Taiwan 10617, Republic of China}


\begin{abstract}
In view of ongoing measurements of the Higgs-like boson at the LHC and direct searches for
dark matter, we explore the possibility of accommodating the potential results in a simple
new-physics model with discrete gauge symmetry as well as light neutrino masses.
Specifically, we study collider and relic-density constraints on the new gauge coupling,
predict the cross section of the dark matter scattering off nucleons, and
compare it with current direct search data.
We also discuss some of the implications if the dark matter is light.
The new gauge sector of the model allows it to be compatible with the latest LHC information on
the Higgs-like particle and simultaneously satisfy the requirements in its dark matter sector.
\end{abstract}

\maketitle

\section{Introduction}
The particle with mass around 125\,GeV recently discovered at the LHC has
characteristics which, based on the measurements reported so far~\cite{lhc},
suggest that it may be the Higgs boson of the standard model (SM).
Whether that is indeed the case will likely become clearer in the near future after
sufficiently more data are accumulated at the~LHC.
Needless to say, the imminent potential confirmation on the Higgs' existence will have
far-reaching implications for attempts to identify the nature of new physics beyond the SM.
For it is widely accepted that new physics is necessary at least to explain the astronomical
evidence for dark matter (DM) and the numerous experimental indications of
neutrino mass~\cite{pdg}.
In anticipating the outcomes of the Higgs quest as well as upcoming results of direct
searches for DM, which are also ongoing, it is of great interest to explore a~simple
framework of new physics that can accommodate the possibilities.

To do so, we adopt an economical model that offers not only a Higgs boson and DM of the popular
weakly interacting massive particle (WIMP) type, but also a means to produce light neutrino masses.
The parameter space of the model has ample room for a Higgs with SM characteristics or one
that is less SM-like.
The Higgs and DM sectors are linked in that the Higgs may decay substantially into DM
particles if kinematically allowed.
This makes the Higgs and DM searches complementary for probing the model.
Another of its salient features is that the DM stability is realized under
a discrete $Z_2$ symmetry that is not imposed in an {\it ad hoc} manner, but instead
emerges naturally as a remnant of an extra Abelian gauge group, U(1)$_\zeta$,
spontaneously broken by the nonzero vacuum expectation value (VEV) of a new scalar
field~\cite{Krauss:1988zc}.
We assume that the SM fermions carry U(1)$_\zeta$ charges, and this allows us to introduce
right-handed neutrinos for gauge-anomaly cancellation and generating light neutrino masses
by means of the well-known seesaw mechanism~\cite{seesaw},
which is activated with the involvement of the same new scalar field.

This paper is organized as follows.
In Sec.\,\ref{model}, we describe our model in greater detail.
We also briefly compare the important features of our model with a number of other scenarios in
the literature.
In Sec.\,\ref{constraint}, we obtain constraints on the new gauge sector from collider experiments.
In Sec.\,\ref{DM}, we address further constraints on the model from the relic abundance data
and predict the DM-nucleon scattering cross-section subject to direct detection measurements.
We also discuss the invisible decay of the Higgs boson if the DM is sufficiently light
and consider some of the implications in relation to the Higgs hunt at the LHC.
Finally, Sec.\,\ref{summary} contains the summary of our work and some additional discussions.

\section{\label{model}Model description}

\begin{table}[b]
\begin{tabular}{|c||c|c|c|c|c|} \hline
~~ & $f_{\rm SM}^{}\vphantom{\int_|^|}$ & $\nu_{kR}^{}$ & $H$ & $S$ & $D$ \\ \hline \hline
\,SU(2), U(1)$_Y\vphantom{\int_|^{|^|}}$\, & $g_{\rm SM}^f$ & 1, 0 & ~2, 1/2~ & 1, 0 & 1, 0 \\ \hline
U(1)$_\zeta~[Z_2]\vphantom{\int_|^{|^|}}$ &
~$\zeta_f$~[$-$]~ & ~$-1$~[$-$]~ & ~0~[+]~ & ~2~[+]~ & ~1~[$-$]~ \\ \hline
\end{tabular} \vspace{-1ex}
\caption{Charge assignments of the fermions and scalars in the model.
$f_{\rm SM}^{}$ $\bigl(g_{\rm SM}^f\bigr)$ denotes SM fermions (their assignments) and
$H$ the usual complex doublet.
For quarks and leptons, \,$\zeta_f=1/3$ and $-1$,\, respectively.\label{contents}} \vspace{-1ex}
\end{table}

We introduce a minimal number of new particles for our purposes: three right-handed neutrinos,
$\nu_{kR}^{}$, two complex scalar fields, $S$ and~$D$, which are singlets under the SM gauge
group, and a $Z'$ boson for the U(1)$_\zeta$ gauge symmetry.
We collect the quantum number assignments for the fermion and scalar
fields of our model in Table~\ref{contents}.
One can therefore see that U(1)$_\zeta$ is none other than U(1)$_{B-L}$~\cite{B-L:Model,b-l},
with $B$ and $L$ referring to baryon and lepton numbers, respectively.
The kinetic terms of $S$ and~$D$ along with~a~renormalizable potential $\cal V$ for
them and $H$ are given by
\begin{eqnarray} \label{DeltaL} && \hspace*{-5ex}
{\cal L}^{} \,\,=\,\, ({\cal D}^\mu D)^\dagger\,{\cal D}_\mu^{}D +
({\cal D}^\mu S)^\dagger\,{\cal D}_\mu^{}S \,-\, {\cal V} ~,
\\ \label{potential}
{\cal V} &=& \mu_D^2 |D|^2 - \mu_S^2 |S|^2 + \mu_{DS}^{} \bigl(D^2 S^\dagger+{\rm H.c}\bigr)
\nonumber \\ && \!\! +\;
2\lambda_{DS}^{}|D|^2|S|^2+2\bigl(\lambda_{DH}^{}|D|^2+\lambda_{HS}^{}|S|^2\bigr)H^\dagger H
\nonumber \\ && \!\! +\; \lambda_D^{}|D|^4 + \lambda_S^{}|S|^4 +
\bigl(\lambda_H^{}H^\dagger H-\mu_H^2\bigr)H^\dagger H  ~, ~~~~~~~
\end{eqnarray}
where \,${\cal D}_\mu=\partial_\mu+i g_\zeta^{}\zeta Z'_\mu$\, with the coupling
$g_\zeta^{}$ and charge~$\zeta$ associated with U(1)$_\zeta$.
The $H$ field develops a~nonzero VEV as in the SM.
The non-SM parts of $\cal V$ were discussed before
({\it e.g.}, Refs.\,\cite{McDonald:1993ex,Batell:2010bp}).
The role of DM is played by the $D$ field---sometimes dubbed the darkon---via its lighter component.
To maintain the $Z_2$ symmetry and hence the DM longevity, $D$ must have zero VEV.
The presence of both the $S$ field and the $\mu_{DS}^{}$ term is essential because the latter
triggers the spontaneous breakdown \,U(1)$_\zeta\to Z_2$\, when $S$ gets a nonzero VEV.

The parameters in $\cal V$ should be chosen such that the vacuum has the above desired
properties.  Accordingly, we assume that all the $\lambda$'s in $\cal V$ are positive
to render it bounded from below.
Subsequently, upon expressing the VEV's of $H$ and $S$ as
\,$\langle H\rangle=(0 ~ ~ v_H^{})^{\rm T}/\sqrt2$\,
and \,$\langle S\rangle=v_S^{}/\sqrt2$,\, with~\,$v_{H,S}^{}>0$,\, we arrive at
\begin{eqnarray}
v_{H(S)}^2 \,\,=\,\, \frac{\lambda_{S(H)\,}^{}\mu_{H(S)}^2-\lambda_{HS\,}^{}\mu_{S(H)}^2}
{\lambda_{H\,}^{}\lambda_S^{}-\lambda_{HS}^2} ~,
\end{eqnarray}
and so
\begin{eqnarray}
\mu_{H(S)}^2 \,\,=\,\, \lambda_{H(S)\,}^{}v_{H(S)}^2+\lambda_{HS\,}^{}v_{S(H)}^2 \,\,>\,\, 0 ~.
\end{eqnarray}
Furthermore, writing \,$D=(D_R+i D_I)/\sqrt2$\, in terms of its real and imaginary components
leads to the combinations
\begin{eqnarray} \label{md}
m_{D_R,D_I}^2 &\!=& \mu_D^2 + \lambda_{DH\,}^{}v_H^2+\lambda_{DS\,}^{}v_S^2 \pm
\sqrt2\, \mu_{DS\,}^{}v_S^{}
\,>\, 0 ~,
\nonumber \\
\end{eqnarray}
which are the squared masses of $D_R$ and $D_I$.
The mass difference between $D_R$ and $D_I$ therefore depends on the sign of $\mu_{DS}^{}$.
We will take \,$\mu_{DS}^{}>0$\, so that $D_I$ acts as the WIMP DM.
The results of our analysis would be the same if we took \,$\mu_{DS}<0$,\,
only that the roles of $D_R$ and $D_I$ would be interchanged.

After electroweak symmetry breaking, the remaining field $h'$  in
\,$H=\bigl(0 ~~~ v_H^{}+h'\bigr){}^{\rm T}/\sqrt2$\, will mix with
\,$s'=\sqrt2\,S-v_S^{}$\, because of the $\lambda_{HS}^{}$ term in~${\cal L}$.
This results in the mass eigenstates
\begin{eqnarray}
h \,\,=\,\, h'\cos\theta+s'\sin\theta ~, ~~~~
s \,\,=\,\, s'\cos\theta-h'\sin\theta ~, ~~
\end{eqnarray}
with the mixing angle $\theta$ and masses $m_{h,s}^{}$ given by
\begin{eqnarray} & \displaystyle
\tan(2\theta) \,\,=\,\, \frac{M_{HS}^2}{M_H^2-M_S^2} ~, & ~~~~
\\ &
2m_{h,s}^2 \,\,=\,\, M_H^2+M_S^2\mp\bigl[\bigl(M_H^2-M_S^2\bigr){}^2+M_{HS}^4\bigr]^{1/2} \,, &
\end{eqnarray}
where
\begin{eqnarray} \label{M2HS}
M_{H,S}^2 \,\,=\,\, 2\lambda_{H,S\,}^{}v_{H,S}^2 ~, ~~~~
M_{HS}^2 \,\,=\,\, 4\lambda_{HS\,}^{}v_{H\,}^{}v_S^{} ~.
\end{eqnarray}
The lighter state, $h$, is the physical Higgs boson.

In the neutrino sector, the mass-generating terms have the form
\begin{eqnarray} \label{numass}
i\lambda_{kl\,}^{}\bar\nu_{kR\,}^{}H^{\rm T}\tau_2^{}L_{lL}^{} \,-\,
\mbox{$\frac{1}{2}$}\lambda_{kl\,}'\bar\nu_{kR\,}^{}(\nu_{lR}^{})^{\rm c}S^\dagger
\;+\; {\rm H.c.} ~,
\end{eqnarray}
where \,$k,l=1,2,3$\, are summed over, $\tau_2^{}$ is the second Pauli matrix, and
$L_{lL}$ represents a lepton doublet.
They give rise to the Dirac and Majorana mass (3$\times$3) matrices
\begin{eqnarray} \label{MD}
{\cal M}_D^{} \,\,=\,\, \mbox{$\frac{1}{\sqrt2}$}\,\lambda\,v_H^{} ~, ~~~~
{\cal M}_{\nu_R}^{} \,\,=\,\, \mbox{$\frac{1}{\sqrt2}$}\,\lambda' v_S^{} ~,
\end{eqnarray}
respectively.
Incidentally, the $\lambda_{kl\,}'$ term in Eq.\,(\ref{numass}) can play the same role
as the $\mu_{DS}^{}$ term in Eq.\,(\ref{potential}) for symmetry breaking.
For the type-I seesaw mechanism to yield the light neutrino masses, the eigenvalues of
${\cal M}_{\nu_R}$ are expected to be orders of magnitude bigger than a TeV~\cite{seesaw}.

Thus, assuming that $\lambda_H^{}$ and $\lambda_S^{}$ in Eq.\,(\ref{M2HS}) are
roughly of similar order to the eigenvalues of $\lambda'$ in Eq.\,(\ref{MD}), we have
\,$M_S^{}\gg1$\,TeV\, and also \,$M_S^{}\gg M_H^{}$.\,
In addition, we will pick \,$M_{HS}^2\ll M_S^2-M_H^2$,\, so that \,$m_{h,s}^{}\simeq M_{H,S}^{}$\,
and~\,$\theta\ll1$,\, leading to \,$h\sim h'$\, and \,$s\sim s'$.\,
It follows that we can neglect the effects of the heavy $s$ on the processes of interest
and the relevant interactions for $D_{R,I}$ are described by
\begin{eqnarray} \label{LD}
{\cal L}_D^{} &=&
-\lambda_{DH}^{}\bigl(D_R^2+D_I^2\bigr)\Bigl(v_H^{}h+\mbox{$\frac{1}{2}$}h^2\Bigr)
\nonumber \\ && -\; \mbox{$\frac{1}{4}$}\lambda_D^{}\bigl(D_R^2+D_I^2\bigr)^2
+ g_\zeta^{}\bigl(D_R^{}\!\stackrel{\leftrightarrow}{\partial}\!\!{}^\mu D_I^{}\bigr)Z_\mu' ~.
~~~~~~~
\end{eqnarray}
From now on, we consider the possibility that $D_{R}$ and $D_I$ are nearly degenerate.
As a consequence, the DM relic abundance is determined not only by the annihilation rate
of~$D_I$, but also by that of $D_R$ and/or both of them.
In this so-called coannihilation case~\cite{Griest:1990kh} we assume specifically that
\begin{eqnarray}
\Delta \,\,=\,\, \bigl(m_{D_R}-m_{D_I}\bigr)/m_{D_I}^{} \,\,\simeq\,\, 0 ~.
\end{eqnarray}
Therefore, the relevant reactions are mainly
\,$D_{I(R)}D_{I(R)}\to\rm SM$ particles,\, from Higgs-exchange and contact diagrams,
the latter if~\,$m_{D_I,D_R}>m_h^{}$,\, and \,$D_I D_R\to Z^{\prime*}\to\rm SM$ fermions,\,
as the fermions carry U(1)$_\zeta$ charges.
Moreover, the $\lambda_D^{}$ part in ${\cal L}_D^{}$ is not pertinent to our purposes.

Before proceeding to our numerical calculations, we would like to make a few remarks
comparing this work to those in the literature.
A number of earlier analyses addressed the phenomenology of some of the elements in our model
separately, such as scalar DM in the absence of a new gauge
sector~\cite{Silveira:1985rk,Davoudiasl:2004be,McDonald:1993ex,He:2010nt,He:2011de}
or the $Z'$ boson of a gauged U(1)$_{B-L}$ symmetry in conjunction with right-handed neutrinos,
but no DM candidates~\cite{B-L:Model}.
There are also studies dealing with models which possess DM stabilized by the remnant
$Z_2$ symmetry of a U(1)$_{B-L}$ that is global~\cite{b-l:global} or local~\cite{b-l,b-l:gauged}.
The $Z_2$ symmetry could instead be simply put in by
hand~\cite{Silveira:1985rk,Davoudiasl:2004be,McDonald:1993ex,He:2010nt,He:2011de,Okada:2010wd},
or it could be accidental~\cite{Gopalakrishna:2009yz}.
Here we would like to emphasize that the model which we have adopted incorporates the minimal
mechanism for both stabilizing scalar DM with the $Z_2$ remnant of a gauged U(1)$_{B-L}$ and
generating neutrino mass through the spontaneous breaking of the same group.
Furthermore, the most nontrivial aspects of our analysis---in distinction to others'---are
that the $Z'$ boson of the U(1)$_{B-L}$ as well as the Higgs boson contribute to the DM
interactions with SM particles, and hence to DM annihilation, and that we treat
the case where the $Z'$ boson plays a dominant role in determining the DM relic density.
As we will demonstrate later, this allows the model to provide not only a good candidate for
DM that is consistent with the latest direct-search results, but also a~Higgs boson that is
SM-like in its couplings to the standard particles and, for sufficiently light~DM,
has a significant invisible decay mode compatible with the current LHC~data.

\section{Constraints on the new gauge coupling\label{constraint}}

Since $\lambda_{DH}$ and $g_\zeta^{}$ in Eq.\,(\ref{LD}) are free parameters,
we will consider different interesting combinations of their contributions to
the relic density and their potential implications for Higgs and DM direct searches.
Before doing so, we first constrain $g_\zeta$ using other observables measured at
colliders.  Since only $S$ induces the $Z'$ mass, \,$m_{Z'}^{}=2g_{\zeta}^{}v_S^{}$,\,
there is no tree-level $Z$-$Z'$ mixing, implying that at tree level $g_\zeta$ has no
effects on the $Z$-pole observables~\cite{Chiang:2011cv}, but can affect $e^+e^-$ and
hadron collisions into fermion pairs.
Neglecting $Z$-$Z'$ kinetic mixing which can arise at loop level, we focus on
\,$e^+e^-\to\ell^+\ell^-$\, and the Drell-Yan (DY) process \,$pp\to\ell^+\ell^- X$,
which can restrict $g_\zeta^{}$ well.

\begin{figure}[t]
\includegraphics[width=85mm]{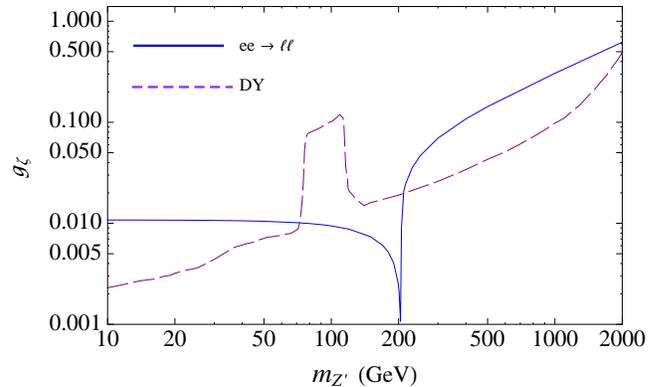} \vspace{-1ex}
\caption{Upper limits on the gauge coupling $g_\zeta^{}$ of the $Z'$ boson versus its mass
$m_{Z'}^{}$ from LEP~II and LHC data on \,$e^+e^-\to\ell^+\ell^-$\, and
\,$pp\to\ell^+\ell^- X$,\, respectively.\label{gzetaLimit}}
\end{figure}

Measurements of \,$e^+e^-\to\ell^+\ell^-$\, for \,$\ell=\mu,\tau$\, were performed
at LEP\,II with center-of-mass energies from 130 to 207~GeV~\cite{Alcaraz:2006mx}.
We employ the data on the cross section and forward-backward asymmetry.
Adopting their 90\% confidence-level (CL) ranges and employing the formulas given
in~Ref.\,\cite{Chiang:2011cv}, but with $s$-dependent $Z$ and $Z'$
widths~\cite{Alcaraz:2006mx}, we extract the upper limit on $g_\zeta$ as a function of
$m_{Z'}^{}$, represented by the blue solid curve in Fig.\,\ref{gzetaLimit}.

The latest cross-section data on the DY process from the LHC~\cite{dy@lhc} reveal no deviation
from the SM expectations and hence no evidence of a $Z'$ boson.
We can derive an upper bound on the coupling constant $g_\zeta$ using the SM cross-section,
following the method of~Ref.\,\cite{Chiang:2011kq}.
The DY cross-section is numerically estimated using the CalcHEP package~\cite{Ref:CalcHEP}
by incorporating new Feynman rules in the model file.  We count the events in
the invariant-mass window of $\pm 20 \%$ around the $Z'$ mass for a luminosity of
$1.1\,{\rm fb}^{-1}$  according to the recent experimental analysis~\cite{dy@lhc}.
This number of signal events is plugged into the one-bin log likelihood
\,$LL=2[N \ln(N/\nu) + \nu - N]$,\,  where $N$ ($\nu$) is the number of events predicted by
the SM (SM plus the $Z'$ boson).
A value of \,$LL=2.7$\, is taken for the $90\%$ CL, and $\nu$ is solved for.
The upper limit on the cross section is then derived from the solved value of $\nu$ for each
$Z'$ mass.
This upper limit in turn constrains $g_\zeta$, as depicted by the purple dashed curve
in Fig.~\ref{gzetaLimit}.

\section{\label{DM} Dark matter phenomenology}

In the simplest darkon model, with a real darkon and no new gauge sector,
the \,$m_D^{}<m_h^{}/2$\, region has mostly been disfavored by DM direct detection results
as well as the recent observation of the Higgs-like resonance at the~LHC, for the darkon-Higgs
coupling is required to be sizable by the relic density data, rendering the Higgs mostly
invisible~\cite{Djouadi:2011aa}.
In our model, the presence of the $Z'$ boson alters the darkon phenomenology in important ways.
Particularly, there are now $Z'$-mediated diagrams, besides the Higgs-mediated ones,
contributing to both darkon annihilation and darkon-nucleon interactions.
One of the consequences is that the allowed parameter space of the model can still
comfortably make room for a~light darkon, as we will see below.

\subsection{Relic abundance}

In evaluating the darkon contributions to the relic density $\Omega_D^{}$,
we employ the relations~\cite{Griest:1990kh,Kolb:1990vq}
\begin{eqnarray} \label{omega} & \displaystyle
\Omega_D^{}h_0^2 \,\,=\,\, \frac{1.07\times10^9}{\sqrt{g_*^{}}\,m_{\rm Pl}^{}\,J\,\rm\,GeV} ~,
& \nonumber \\ & \displaystyle
J  \,\,=\,\, \raisebox{0.7ex}{\footnotesize$\displaystyle\int_{x_f}^\infty$} dx\;
\frac{\bigl\langle\sigma_{\rm eff}^{}\,v_{\rm rel}^{}\bigr\rangle}{x^2} ~,
& \nonumber \\ & \displaystyle
x_f^{} \,\,=\,\, \ln\Bigl[0.038\,g_{\rm eff}^{}\,m_D^{}\,m_{\rm Pl}^{}\,
\bigl\langle\sigma_{\rm eff}^{}v_{\rm rel}^{}\bigr\rangle\bigl(g_*^{}x_f^{}\bigr){}^{\!-1/2}
\Bigr] \,, & ~~~~~
\end{eqnarray}
where $h_0$ denotes the Hubble constant in units of 100\,km/s/Mpc,\,
$g_*^{}$ is the number of relativistic degrees of freedom below the freeze-out
temperature~$T_f^{}$, \,$m_{\rm Pl}^{}=1.22\times10^{19}$\,GeV\,~is the Planck mass,
\,$x_f^{}=m_D^{}/T_f^{}$\,~with \,$m_D^{}=m_{D_I}$~being the WIMP mass,
\,$\langle\sigma_{\rm eff}v_{\rm rel}\rangle\equiv\langle\sigma v\rangle$\, is the thermally
averaged product of the effective darkon-coannihilation cross-section and the relative speed of
the darkon pair in their center-of-mass frame, and $g_{\rm eff}^{}$ is the darkon's effective
number of degrees of freedom in the coannihilation case.
Our choice \,$\Delta\simeq0$\, above leads to some simplification~\cite{Griest:1990kh}.
Thus, we have \,$g_{\rm eff}^{}\simeq2$\, and
\begin{eqnarray}
\sigma_{\rm eff}^{} \,\,\simeq\,\, \mbox{$\frac{1}{4}$}
\bigl( \sigma_{II}^{} + 2\sigma_{IR}^{} + \sigma_{RR}^{} \bigr) ~,
\end{eqnarray}
where $\sigma_{ij}^{}$ denotes the cross section of $D_i D_j$ annihilation into possible
SM final states.  Since $D_{I,R}$ have the same interactions [Eq.\,(\ref{LD})],
we have \,$\sigma_{RR}^{}\simeq\sigma_{II}^{}$\, for \,$\Delta\simeq0$.\,

The expression for $\sigma_{II}^{}$ due to Higgs-exchange diagrams follows from its
real darkon counterpart~\cite{Silveira:1985rk,He:2010nt}, and so
\begin{eqnarray} \label{csii}
\bigl\langle\sigma_{II}^{} v_{\rm rel}^{}\bigr\rangle \,\,=\,\,
\frac{4\lambda_{DH\,}^2 v_H^2\,m_D^{-1}\,\mbox{$\sum_i$}\Gamma\bigl(\tilde h\to X_i\bigr)}
{\bigl(4m_D^2-m_h^2\bigr){}^{^{\scriptstyle2}}+\Gamma^2_h\,m^2_h} ~,
\end{eqnarray}
where $\tilde h$~is a virtual Higgs with the same couplings as the physical $h$,
but with the invariant mass \,$\sqrt s=2m_D^{}$, and  \,$\tilde h\to X_i$\, is any
kinematically allowed decay mode of~$\tilde h$.
For~\,$m_D^{}>m_h^{}$,\, the \,$D_I D_I\to hh$\, channel needs to be included~\cite{He:2010nt}.
Thus the contributions of $\sigma_{II,RR}^{}$ to \,$\langle\sigma v\rangle$\, and $J$ are
\begin{eqnarray}
\langle\sigma v\rangle_h^{} \,\,\simeq\,\,
\frac{\bigl\langle\sigma_{II}^{} v_{\rm rel}^{}\bigr\rangle}{2} ~, \hspace{5ex}
J_h^{} \,\,\simeq \,\,\frac{\langle\sigma_{II}^{}v_{\rm rel}^{}\rangle}{2 x_f^{}} ~.
\end{eqnarray}
For the $Z'$-mediated counterpart, assuming nonrelativistic darkons
and  \,$\Delta\simeq0$,\, we derive
\begin{eqnarray} \label{csIR}
\sigma_{IR}^{} \,\simeq\, \frac{g_\zeta^4 v_{\rm rel}^{}}{12\pi}
\raisebox{0.7ex}{\footnotesize$\displaystyle\sum_f$}
\frac{\zeta_f^2 N_{\rm c}^f\bigl(2m_D^2+m_f^2\bigr)\sqrt{1-m_f^2/m_D^2}}
{\bigl(4m_D^2-m_{Z'}^2\bigr){}^{^{\scriptstyle2}}+\Gamma_{Z'}^2m_{Z'}^2} ~, ~
\end{eqnarray}
where the sum is over fermions with mass \,$m_f^{}<m_D^{}$\, and $N_{\rm c}^f$ colors
and $\Gamma_{Z'}$ is the $Z'$ width.
It follows that
\begin{eqnarray}
\langle\sigma v\rangle_{Z'}^{} \,\,\simeq\,\, \frac{3b_{IR}^{}}{x} ~, \hspace{5ex}
J_{Z'}^{} \,\,\simeq\,\, \frac{3 b_{IR}^{}}{2 x_f^2} ~,
\end{eqnarray}
where \,$b_{IR}^{}=\sigma_{IR}^{}/v_{\rm rel}^{}$.\,
Applying
\begin{eqnarray} \label{svj}
\langle\sigma v\rangle \,\,=\,\,
\langle\sigma v\rangle{}_h^{}+\langle\sigma v\rangle{}_{Z'}^{} ~, \hspace{5ex}
J \,\,=\,\, J_h^{}+J_{Z'}^{}
\end{eqnarray}
in Eq.\,(\ref{omega}), we can then extract constraints on $\lambda_{DH}^{}$ and
$g_\zeta^{}$ from the 90\%-CL range~\,$0.092\le\Omega_D^{}h^2\le0.118$\,
of the observed relic density~\cite{wmap7}.

\begin{figure}[t]
\includegraphics[width=85mm]{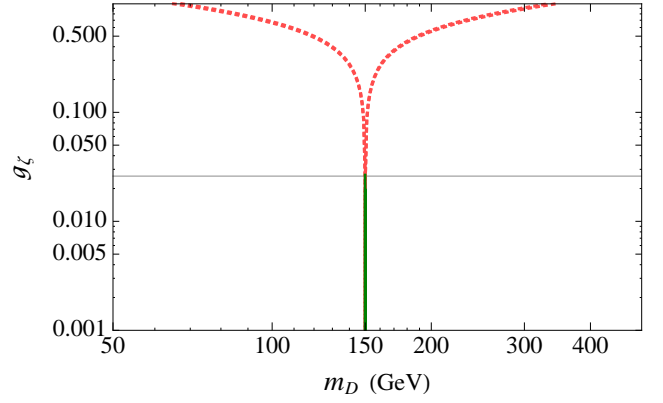} \vspace{-1ex}
\caption{Curve of $g_\zeta^{}$ versus $m_D^{}$ subject to the relic abundance data
for \,$m_{Z'}^{}=300$\,GeV\, and no Higgs-mediated contribution.  The green solid
(red dotted) section is allowed (excluded) by collider data.\label{DMrelic}}
\end{figure}

In Fig.~\ref{DMrelic}, we display $g_\zeta^{}$ as a function of $m_D^{}$ subject to
the relic data, assuming the absence of the Higgs effect and, for definiteness,
\,$m_{Z'}^{}=300$\,GeV,\, and compare it with the collider bound.
This illustrates that only the resonance region, \,$m_{Z'}^{}\simeq 2m_D^{}$,\, can
satisfy both sets of constraints.
Since the situation is unchanged in the presence of the Higgs contribution,
hereafter we limit the $Z'$ contribution to this resonance case.
If the former is nonnegligible, we can consider various combinations of $J_h$ and $J_{Z'}$
with different implications for the prediction in relation to DM direct searches.

\subsection{Direct detection}

The direct detection of DM is through the recoil of nuclei after it hits a nucleon~$N$.
In our model, with nearly degenerate $D_{I,R}$, the involved interactions are
\,$D_{I(R)}N\rightarrow D_{I(R)}N$\, and \,$D_{I(R)}N\rightarrow D_{R(I)}N$\,
via $h$ and $Z'$ exchanges, respectively, in the $t$ channel.
The resulting spin-independent cross-section $\sigma_{DN}^{}$ needs to accommodate these
possibilities, but also take into account the fact that the DM local density is independent
of the number of DM components.  Accordingly, we find
\begin{eqnarray} \label{csDN}
\sigma_{DN}^{} \,\,\simeq\,\,
\frac{\lambda_{DH\,}^2g_{NNh\,}^2\mu_{DN\,}^2v_H^2}{\pi\,m_{D\,}^2m_h^4}
\,+\, \frac{g_{\zeta\,}^4\mu_{DN}^2}{\pi\,m_{Z'}^4}
\end{eqnarray}
in the nonrelativistic limit, where $g_{NNh}^{}$ is the Higgs-nucleon effective coupling,
\,$\mu_{DN}^{}=m_D^{}m_N^{}/\bigl(m_D^{}+m_N^{}\bigr)$,\, the first term is equal in form to
the cross section in the real darkon case~\cite{Silveira:1985rk,He:2010nt}, and
for the second term we have used
\,$\langle N|\bar u\gamma^\alpha u+\bar d\gamma^\alpha d|N\rangle=3\bar N\gamma^\alpha N$\,
with vanishing contributions from the other quarks~\cite{Kaplan:1988ku}.

\begin{figure}[t]
\includegraphics[width=85mm]{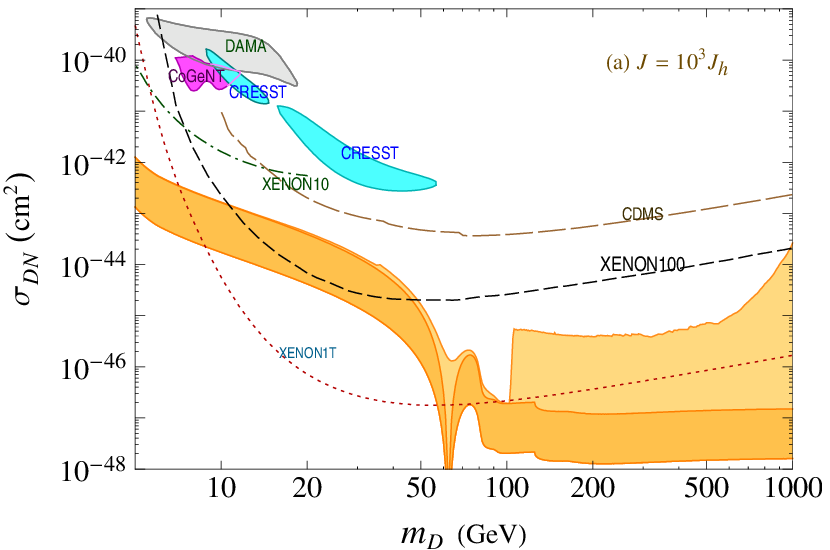}\\
\includegraphics[width=85mm]{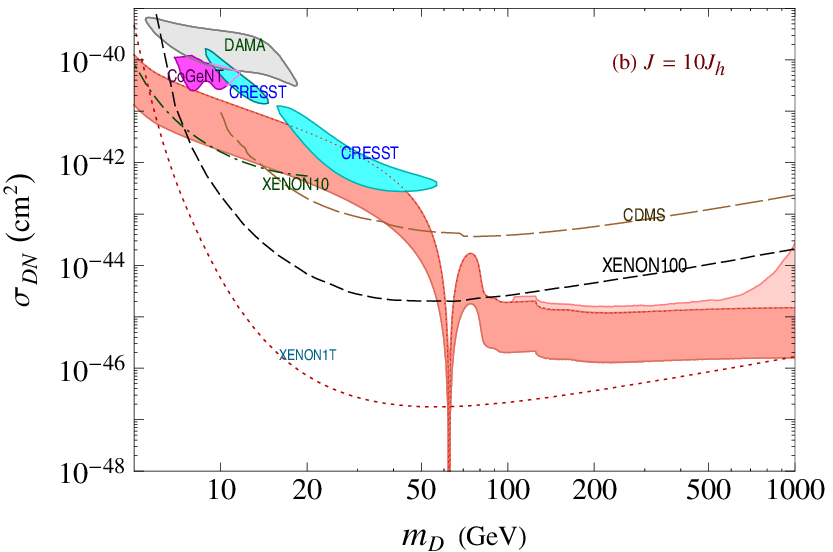}\vspace{-1ex}
\caption{Darkon-nucleon scattering cross-section corresponding to (a) $J=10^3J_h$
(orange regions) and (b) $J=10J_h$ (red regions)\, for \,$m_h^{}=125$\,GeV\,
and \,$m_{Z'}^{}=2m_D^{}$.
The darker (lighter) portions of the predictions come from the contributions of $h$ alone
(both $h$ and $Z'$).
The predictions are compared to {90\%-CL} upper limits from CDMS, XENON10, and
XENON100~\cite{cdms}, as well as the {90\%-CL} signal (purple) region suggested by CoGeNT,
a (gray) patch compatible with DAMA modulation signal at the 5$\sigma$ level,
and two 2$\sigma$-confidence (blue) areas representing CRESST-II data~\cite{cogent}.
Also plotted is the XENON1T projected sensitivity~\cite{xenon1t}.\label{DN2DN}}
\end{figure}

Here we look at a~couple of representative examples.
In the first one, the $Z'$ contribution considerably dominates $J$ in Eq.\,(\ref{svj}),
and we choose \,$J_{Z'}=999\,J_h$\, for definiteness.
We illustrate the prediction for $\sigma_{DN}^{}$ as a function of the darkon mass
with \,$m_h^{}=125$\,GeV\, and~\,$m_{Z'}^{}=2m_D^{}$\, in~Fig.\,\ref{DN2DN}(a).
For this plot, $\lambda_{DH}$ follows from \,$J=10^3J_h$\, plus the relic
constraint, $g_\zeta^{}$ varies in the range allowed by the collider data,
\,$v_H^{}=246$\,GeV,\, and we have employed the range \,$0.0011\le g_{NNh}^{}\le0.0032$\,
for the Higgs-nucleon coupling~\cite{He:2011de,Cheng:2012qr}.
The lightly shaded (lighter orange) portions of the prediction curve indicate that in
the \,$m_D^{}\mbox{\small\;$\gtrsim$\;}100$\,GeV\, region the $Z'$ effect also dominates
$\sigma_{DN}^{}$ and is enhanced by a few orders of magnitude relative to the Higgs
contribution.
Evidently, compared to the most recent data from the leading direct searches for WIMP DM,
the prediction can largely escape the strictest bounds to date.
However, future direct searches such as XENON1T would probe this parameter space of
the model more stringently.

In Fig.\,\ref{DN2DN}(b), we show an example where the Higgs contribution is
less suppressed, \,$J_h=J_{Z'}/9$.\,
In this case, most of the \,$m_D^{}<100$\,GeV\, range is ruled out by the null results of
some of the direct searches.
Nevertheless, the prediction (red areas) is partly consistent with the possible WIMP hints
reported by CoGeNT (purple areas) and CRESST-II (blue areas), although they conflict with
the other experiments.

If the darkon annihilation is dominated by the Higgs contribution instead, \,$J_h\gg J_{Z'}$,\,
the allowed parameter space for \,$m_D^{}<100$\,GeV\, would be further reduced compared to
that in the second example.
With \,$J\simeq J_h$,\, the relic density would comprise approximately equal parts
from $D_{I,R}$ and consequently $\lambda_{DH}^{}$ $\bigl(\sigma_{DN}^{}\bigr)$ would be about
$\sqrt2$~(2) times the corresponding coupling (cross section) in the simplest darkon model.
We note that in all these instances darkon masses larger than \,{\small$\sim$}100\,GeV\,
are still viable and will be probed by future measurements.

\subsection{Invisible decay of the Higgs boson}

\begin{figure}[b]
\includegraphics[width=85mm]{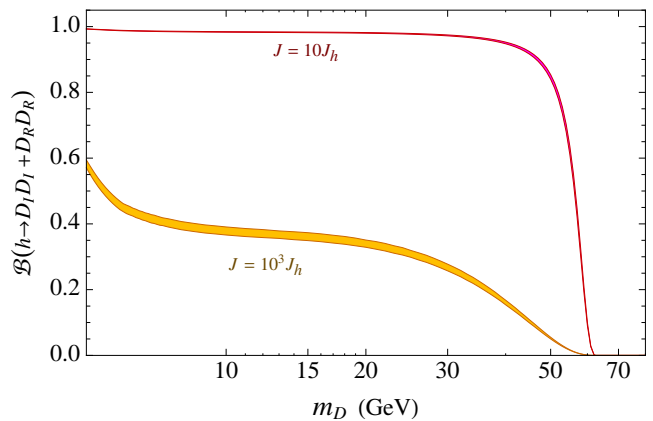} \vspace{-1ex}
\caption{Total branching ratios of \,$h\to D_{I,R} D_{I,R}$\, versus $m_D^{}$ corresponding
to \,$J_{}=10J_h$ and $10^3J_h$\, for \,$m_h^{}=125$\,GeV.\label{hBR}}
\end{figure}

For \,$m_h^{}>2m_{D_R,D_I}$,\, the decays \,$h\to D_{R,I}D_{R,I}$\, will occur and, for
non-negligible $\lambda_{DH}$, can substantially enhance the invisible decay rate of the Higgs.
Their combined branching ratio is
\begin{eqnarray}
{\cal B}(h\to D_I D_I+D_R D_R) \,=\,
\frac{\Gamma_{h\to D_I D_I}+\Gamma_{h\to D_R D_R}}{\Gamma_h} ~~~
\end{eqnarray}
where the Higgs width \,$\Gamma_h=\Gamma_h^{\rm SM}+\Gamma_{h\to D_I D_I}+\Gamma_{h\to D_R D_R}$\,
includes the SM Higgs width $\Gamma_h^{\rm SM}$.
We depict in Fig.\,\ref{hBR} the branching ratios for the same $J_{h,Z'}$
and $m_h^{}$ choices made in Fig.\,\ref{DN2DN} and \,$\Delta\simeq0$.\,
Clearly, unless $J_h$ is very small relative to~$J_{Z'}$, these invisible channels tend
to dominate the Higgs width.
This is similar to the real darkon case~\cite{Silveira:1985rk,He:2010nt,He:2011de},
in which the Higgs can be hidden from sight for sufficiently low \,$m_h^{}>2m_D^{}$.\,
The model can thus readily account for the possibility that the LHC does not detect any Higgs.
On the other hand, if an SM-like Higgs is observed at {\small$\sim$}125\,GeV,
the model can also accommodate it and still provides light DM with the right relic abundance
via the $Z'$ contribution.
In contrast, the simplest real-darkon model with \,$m_D^{}<m_h^{}/2$\, would be disfavored by
such a discovery~\cite{Djouadi:2011aa}.
More generally, depending on the parameters, our model can explain a partially hidden Higgs,
while leaving the ratios of branching fractions to SM particles essentially the same.
For \,$2m_{D_I,D_R}>m_h^{}$,\, the Higgs decay pattern would be SM-like.

The recently discovered particle of mass near 125\,GeV has properties consistent with
those of an SM Higgs, based on the LHC information reported so far~\cite{lhc}.
However, the present data still allow the new boson to have a branching ratio into invisible
particles of up to a few tens of percent~\cite{h2inv}.
All this is compatible with one of the possibilities discussed above, where
the $Z'$ contribution greatly dominates the relic density.
Obviously, future measurements at the LHC and DM direct searches together will test such
a~scenario within our model.

\section{Summary and discussion\label{summary}}

Anticipating the upcoming results of the Higgs hunt at the LHC and direct searches for DM
in various underground experiments, we have considered a~simple model possessing
only a~small number of nonstandard particles, including
scalar DM---the darkon---which is stabilized by a~$Z_2$ symmetry naturally arising from
the spontaneous breaking of a~gauged U(1)$_{B-L}$ group.
The associated gauge boson,~$Z'$, yields new effects on the DM relic
abundance in the resonant case, \,$m_{Z'}^{}\simeq 2m_D^{}$.\,
We use the measurements of \,$e^+e^-\to\ell^+\ell^-$\, and \,$pp\to\ell^+\ell^-X$
($\ell = \mu,\tau$) and the relic density data to constrain the new gauge coupling.
Subsequently, we explore different combinations of the Higgs- and $Z'$-mediated contributions
to darkon annihilation subject to the relic density constraints and evaluate the corresponding
darkon-nucleon scattering cross-section, $\sigma_{DN}^{}$, compared to direct search results.
If the $Z'$ effect dominates the darkon relic density,  $\sigma_{DN}^{}$ tends to evade
the present limits and may come mainly from the $Z'$-mediated contribution, depending on
the darkon mass.
We also discuss some implications of the invisible decay of the Higgs if the darkon is
sufficiently light,~\,$m_D^{}<m_h^{}/2$.\,
In that case, we find that the Higgs invisible decay branching-fraction is still significant
for most of the allowed $m_D^{}$ values even if the Higgs contribution to darkon annihilation
is very small compared to the $Z'$ contribution.
This result is important because the invisible branching fraction of the Higgs-like particle
recently observed at the LHC has now been estimated from current data to reach up to
a few tens of percent, which implies, within the context of our model, the necessity of a highly
dominant $Z'$ effect on the DM relic density.
In general, the parameter space of the model has enough room to accommodate various potential
outcomes of the ongoing Higgs and DM direct searches, which will therefore probe the model further.
The LHC can offer additional tests via processes that can produce the $Z'$ and/or a~darkon pair,
the latter due to diagrams mediated by $h$ or $Z'$.
The $Z'$ may be most detectable in final states containing a pair of charged leptons,
whereas the darkon may be uncovered in events with missing energy.

\acknowledgments
This research was supported in part by the National Science Council of Taiwan, R. O. C.,
under Grants No. NSC-100-2628-M-008-003-MY4 and No. NSC-100-2811-M-002-090, and by the
NCU Plan to Develop First-Class Universities and Top-Level Research Centers.

\end{document}